\documentclass[aps,prl,preprint,showpacs,floatfix]{revtex4}
\usepackage{graphicx}

\begin{document}

\title{Charge Carrier  Transport in Metal Phthalocyanine Based Disordered Thin Films } 

\author{Ajit Kumar Mahapatro and 
Subhasis Ghosh}

\affiliation{School of Physical Sciences, Jawaharlal Nehru University,
New Delhi 110067, India}

\begin{abstract}

The  charge carrier  transport in metal phthallocyanine based disordered thin films  has been investigated.   Charge carrier  mobility  in these disordered thin films strongly depends on the  electric field and temperature due to hopping conduction. The applicability of two different Gaussian disorder models has been  compared and evaluated   for charge carrier transport using  simple experimental results and based on our extensive analysis, it has been found  that spatial and energetic correlation is important in explaining  the electrical transport in these organic semiconductors.

\end{abstract}

\pacs{72.80.Le, 72.15.Cz, 72.20.Ee}

\maketitle

The discovery and subsequent  success  of organic light emitting diodes(OLED) based on small
organic molecule aluminum tris-(8-hydroxyquinoline)(Alq3)\cite{cwt87} and polymer
poly(p-phenylene vinylene)(PPV)\cite{jhb90}  lead to tremendous
amount of work devoted in understanding optical and transport
properties of organic semiconductors(OS).    Several models for current transport mechanism, each starting from
different assumptions, have been proposed to explain
the experimental results. Moreover, the  transport processes in
these systems are often influenced by the
peculiarities of the individual systems.  Systematic investigation of both  aspects of current conduction mechanism, i.e.  the charge injection
 from metal electrode to OS and bulk charge transport in OS   have been
reported in polymer-based OS\cite{ihc98} and  molecular OS like,  Alq3\cite{ihc99}. Recently we have  reported\cite{akm02} a detailed study of the charge  injection in metal phthallocyanine(MePc) based  devices,  but so far, now more systematic investigations  on bulk charge  transport are required in this technologically important OS,  which has
potential\cite{zb96,gp98,sck01,jz04} to revolutionanize the display technology by enabling
low cost, large area full color flat panel display. MePcs are
macrocyclic metal complexes and have attracted much interest
because of their high chemical stability, various synthetic
modifications, ease to prepare the devices and reproducibility of experimental data, which is a major problem with most other OS like, Alq3\cite{dk01}  due to their degradation with time.

There are several distinguishing features of these disordered solids, (i) they are composed of organic molecules, held together loosely by   weak van der Waals type intermolecular coupling while the intramolecular coupling is strong; (ii) the absence of long range order in these disordered materials lead to the localization of the electronic wave function and the formation of a broad Gaussian density of states(GDOS) and  the most important one is (iii) the carrier mobility $\mu$ exhibits a nearly universal Poole-Frenkel(PF) behavior $\mu(E,T)=\mu(0,T) \exp\left(\gamma(0,T) \sqrt{E}\right)$,  where  $\mu(0,T)$ and $\gamma(0,T)$ are temperature dependent quantities, known as the zero field mobility and the field activation of the mobility, respectively.  In disordered organic  thin films,  hopping among the molecular sites having comparable energies, describes the transport of charge carriers through the GDOS of highest occupied molecular orbital(HOMO) and/or lowest unoccupied molecular orbital(LUMO). In this paper, we present  the experimental investigation on charge carrier  transport in hole only 
devices based on metal/MePc/metal
structures.   The temperature dependence of carrier mobility has been determined from the current-voltage(J-V) characteristics under the space charge limited(SCL) conduction regime, which can be observed when one contact injects more carriers than the two terminal device has in equilibrium\cite{mp99}. This method is preferred over transient based experiment like time-of-flight(TOF)  measurements due to its intrinsic problem in data  interpretation at low temperature and  severe constraints on the sample geometry\cite{pwm98,sb02}.   

We have chosen two different MePcs, copper phthalocyanine(CuPc) and zinc phthalocyanine(ZnPc) for our investigations. Thermally evaporated  thin films of high pure MePcs, procured from Aldrich Co.  are characterized using atomic force microscopy(AFM) and absorption spectroscopy. Details about the MePc-based  metal/MePc/metal devices are given in Ref.\cite{akm02}. Fig.1  shows the AFM images of 100nm,  200nm and 400nm thick CuPC layers grown at room temperature. Surface morphology in both cases indicates amorphous film. It has been shown by Bao et al\cite{zb96} that only elevated substrate temperature($\geq$85$^o$C)  results  polycrystalline or ordered CuPc thin films. We have also observed that grainy structures develop in thicker($\geq$300nm)  CuPc films grown at room temperature.  To investigate  the current transport mechanism in disordered molecular thin films, we have avoided (i) elevated substrate temperature and (ii) higher thicknesses.   Fig.2 and Fig.3 also   show the J-V characteristics of ITO/CuPc/Al
and ITO/CuPc/Cu at room temperature for two thicknesses(100nm and 200nm) of CuPc. The experiment consisted of
two steps. First: the current due to hole injection from positively biased ITO was measured 
and second: the current due to hole injection from Cu and Al was measured
by reversing the polarity of the bias voltage(i.e., biasing Al and Cu electrodes positively). It is clear from  Fig.2 and Fig.3
that J-V characteristics in case of ITO/MePc/Cu display almost symmetric
behavior\cite{akm02} in both cases(hole injection either from ITO, or from Cu
electrodes). This can be explained by a  small energy
barrier of 0.1eV at  ITO/MePc  and 0.2eV in  Cu/MePc interfaces, giving rise to SCL bulk  current when either ITO or Cu is positively biased. This is  because the ionization potential of MePc is 4.8eV\cite{gp98} and the work function of ITO and Cu are 4.7eV and 4.6eV, respectively. It has been shown\cite{akm02,psd97}, for Schottky energy barrier(SEB) less than about 0.3-0.4eV, the current transport is due to SCL. 
 But, in case of ITO/MePc/Al devices, J-V characteristics displays asymmetric and rectification-like
behavior\cite{akm02}. A schematic energy level diagram of the different single layer devices using different contacting metals, used in this investigation is shown in Fig.4. Current density increases by almost five order of
magnitude by reversing the polarity of the bias. As discussed before,
when ITO is positively biased, current is due to SCL and when Al
is positively biased, current is due to injection limited and
reduced by several order of magnitude due to the existence of SEB
of 0.6eV at MePc/Al contact. Similar J-V characteristics are observed in single layer devices with ZnPc. It has been shown by us\cite{akm02} that this injection limited current $J_{inj}$ from a metal electrode to  disordered OS can be described by the relation\cite{ys01},  $J_{inj}\propto\mu\psi^2 exp \left (-\frac{\phi_B}{k_BT} \right )exp \left (\sqrt{\frac{e^3E}{2\pi\epsilon(k_BT)^2}}\right)$, where $\mu$ is carrier mobility, $\psi$ is a slow varying function of electric field $E$,  $\phi_B$ is SEB and $\epsilon$ is dielectric constant. Fig.5 shows the temperature dependent SCL current in  ITO/CuPc/Al. At low bias, J-V characteristic follow the slope of 2 in log(J)-log(V) plot, but as the bias increases the slope   increases gradually to higher value. A numerical workout has been  used to simulate the experimental data by simultaneously solving  Poission's equation, $dE(x)/dx=ep(x)/\epsilon$,  $J(x)=ep(x)\mu(x)E(x)$ and  Poole-Frenkel(PF) relation, $\mu(E,T)=\mu(0,T) \exp\left(\gamma(0,T) \sqrt{E}\right)$.  Here,  $p(x)$ is the hole density at position $x$ and  the boundary condition is taken  at the ITO/MePc interface with  hole density(at $x=0$) of   $2.5\times 10^{19} cm^{-3}$, which is the density of states of highest occupied molecular orbital(HOMO)  in MePc\cite{rac86,aa91}. The simulated results are shown as solid lines in  Fig.5 at different temperatures.  The excellent  agreement of the simulated and the experimental data confirms PF behavior of carrier  mobility.  It has been shown\cite{ct03,ct04,pwmb05} that in case of lower thickness and low carrier concentration, hole mobility depends only on electric field and temperature, but in case of higher thickness($>$200nm) and high carrier concentration($>10^{16}cm^{-3}$), in addition to electric field and temperature, hole mobility also depends on charge carrier concentrations.

Now, we compare our experimental results with two    models for charge carrier transport in disordered molecular solids. Those are  (a) uncorrelated Gaussian disorder model (UGDM) and (b) correlated Gaussian disorder model(CGDM), for  bulk current transport in disordered OS. 
At every temperature,  $\mu(0,T)$ and  $\gamma(0,T)$ are obtained from the experimental data using these two models and plotted in Fig.6.
 Bassler proposed\cite{hb93} the UGDM and provided support to PF behavior of mobility using Monte Carlo simulation. The UGDM describes the carrier transport as a biased random walk among the hopping sites  with Gaussian-distributed random site energies.  In this case, $\mu(0,T)$ and $\gamma(0,T)$ have $1/T^2$ dependence and  can be expressed as
$\mu(0,T) =\mu_{0}\exp \left(-\frac{2}{3}\left (\frac{\sigma}{k_BT}\right )^2\right),
\gamma(0,T)=C\left( \left(\frac{\sigma}{k_BT}\right )^2-\sum^2\right)$, where $\sigma$ is the width of the Gaussian distribution of hopping sites in HOMO, C and $\sum$ are the parameters of the model. The $\mu(0,T)$ and $\gamma(0,T)$,  determined from  fit to
experimental J-V characteristics are plotted in Fig.6a  and the different parameters determined using this model  are given in   Table.I.  Although UGDM explains some features of experimental data,  several discrepancies emerge with uncorrelated description of Gaussian disorder model, which will be discussed later.  The most important criticism against UGDM is its' inability to reproduce the PF behavior over wider range of electric field emphasizing\cite{yng95} the importance of spatial correlation  among the hopping sites for the description of PF behavior of mobility  for wider range of electric field.  It has been shown\cite{dhd96,snn98}   that the interaction of charge carriers with permanent dipoles located on either dopant or host molecules gives rise to spatial and energetic  correlation  and provides physical origin behind universally observed PF behavior of mobility. Essentially, CGDM is based on long-range correlation between charge carriers and the molecular electric dipole, resulting  random potential energy landscape with long-range spatial correlations
$<U(0)U(r)> \sim \sigma^2a/r$,  where $a$ is the minimal charge-dipole separation or the distance between two hopping sites.  In this case,  $\mu(0,T)$ has a $1/T^2$ dependence(same as in UGDM) but $\gamma(0,T)$ has a $1/T^{3/2}$ dependence.  According to this model,  $\mu(0,T)$ and $\gamma(0,T)$ can be expressed as, $\mu(0,T)=\mu_{0} exp\left(-\frac{3}{5}\left (\frac{\sigma}{k_BT}\right )^2\right), 
\gamma(0,T)=C_0 \left( \left (\frac{\sigma}{k_BT}\right )^{3/2}-\Gamma \right)\sqrt{\frac{ea}{\sigma}}$, where C$_0$,  $A$ and $\Gamma$ are parameters of the model. The $\mu(0,T)$ and $\gamma(0,T)$,  determined from  fit to
experimental J-V characteristics are plotted in Fig.6b  and the different parameters determined using this model  are given in   Table II.

The main feature of UGDM is the non-Arrehenious behavior of the temperature dependence of mobility, which is the consequence of hopping conduction in GDOS. But, there are several discrepancies, which are,  (i) fitting is not as good as in case of CGDM, (ii) {\sl the intercept of $\gamma(0,T)$ vs. 1/T$^2$ gives positive value, but according to UGDM relation  it should be negative, which is a fundamental problem with UGDM}, (iii) it has been observed that there is no specific trend in the values of $C$, which is scaling parameter in the model and cannot be linked to a physical parameter of the system.
By  comparing our experimental data  with CGDM,  we observe, (i) the linear dependence of ln$\mu(0,T)$ on $1/T^2$ and $\gamma(0,T)$ on $1/T^{3/2}$ is evident in Fig.6b  and straight line fit is excellent and better than that in UGDM, (ii) {\sl for all samples the intercept of $\gamma(0,T)$ vs. $1/T^{3/2}$ gives negative values},  (iii) the experimental value of $\Gamma$ are 1.2 in CuPc and 1.4 in ZnPc, which are very close to the predicted value 1.97\cite{snn98}, (iv) the width of the Gaussian distribution $\sigma$ is 110meV in CuPc and  120meV in  ZnPc and similar values for $\sigma$ are commonly observed in other OSs.

The UCDM and CGDM share the common feature  regarding   $\mu(0,T)$ due to similar distribution of hopping sites energies and the obtained values match well with the values obtained by independent measurements\cite{zb96,jz04}, but  $\gamma(0,T)$ decides critically the importance of spatial correlation and decides the mechanism of carrier transport in OS.
 In CGDM, it has been shown that long-range interaction between charge carriers and permanent dipole moments of doped molecules in polymers and  host molecules leads to spatial correlation. However, it has been pointed out by Yu {\sl et. al.}\cite{zgy00} that the mechanism responsible for PF behavior in different conjugated polymers and molecules cannot be due to only charge-dipole interaction, because  PF behavior has been universally observed in several doped and undoped polymers and molecules with or without permanent dipole moment.  Hence, in addition to charge-dipole interaction there may be another mechanism responsible for spatial correlation, which is a fundamental requirement for PF behavior. Yu {\sl et. al.}\cite{zgy00} have  shown using first principle quantum chemical calculation that the thermal fluctuations in the molecular geometry can lead to spatial correlation. Identification of exact origin of spatial correlation in these OS based thin films would be an interesting future topic.

In conclusion, we have  presented a comparison of the transport properties with uncorrelated and correlated disorder models in MePc. We have shown  temperature  dependence of different transport parameters can be fitted and described within the correlated Gaussian disorder model, signifying the important role of  correlation among disorders on the carrier transport in these materials. The origin of correlation has been discussed. We hope present analysis would lead to further experimental and theoretical investigations focusing the exact origin and role of correlation on transport properties of technologically important OSs.

\newpage

\noindent {\large \bf Table Captions}

\vspace{0.2in}

\begin{description}

\item[Table I.] Various parameters evaluated using  UGDM.  $\mu_0$ and  $\sigma$  and  scaling parameters $C$ and  $\Sigma$ determined from the ln$\mu(0)$ vs. $1/T^2$ and $\gamma$ vs. $1/T^2$ plots(Fig.5a), respectively,    for CuPc and ZnPc.

\item[Table II.] Various parameters evaluated using  CGDM.  $\mu_0$ and  $\sigma$  and  scaling parameters  $C_0$ and $a$, determined from the ln$\mu(0)$ vs. $1/T^2$ and $\gamma$ vs. $1/T^{3/2}$ plots(Fig.5b), respectively, for CuPc and ZnPc.

\end{description}

\newpage

\begin{center}
Table I
\end{center}
\begin{center}
\begin{tabular}{|ccccc|}\hline 
 MePc & $\mu_0$ & $\sigma$ & C & $\Sigma$\\
 & in $cm^2/Vs)$ & in meV & &  \\  \hline
CuPc	 & $2.8\times 10^{-8}$ & 100 & $6.4\times 10^{-5}$ & -3.1\\	
ZnPc	 & $1.7\times 10^{-8}$ & 110 & $ 4.5\times 10^{-5}$ & -4.24\\  \hline
	\end{tabular}
\end{center}

\newpage

\begin{center}
Table II
\end{center}
\begin{center}
	\begin{tabular}{|ccccc|} \hline 
 MePc & $\mu_0$ & $\sigma$ & C$_0$ & a \\ 
	 & in $cm^2/Vs$ & in meV & & in $\AA$ \\	 \hline
CuPc	 & $2.8\times 10^{-8}$ & 110 & $6.4\times 10^{-5}$ & 20 \\		
ZnPc	 & $1.7\times 10^{-8}$ & 120 & $4.5\times 10^{-5}$ &17 \\		\hline		
	\end{tabular}
\end{center}

\newpage

\noindent {\large \bf Figure Captions}

\vspace{0.2in}

\begin{description}

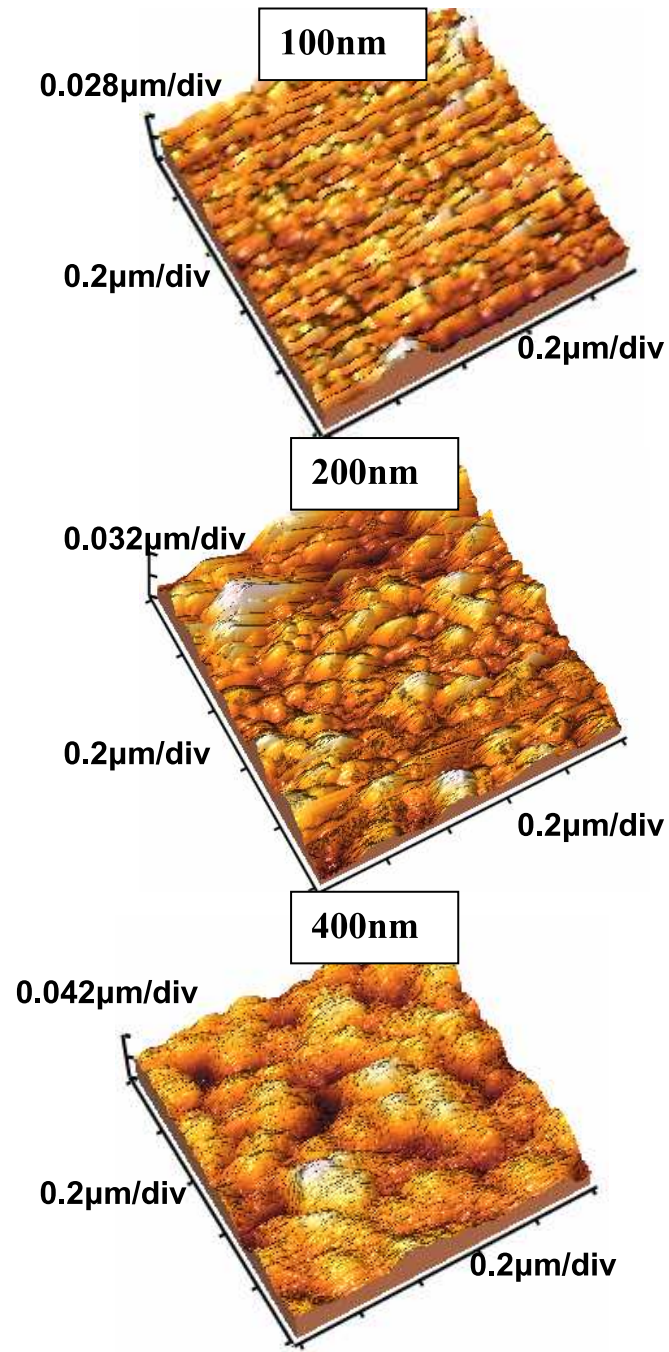
\item[Fig.1] Surface morphology(by AFM)  of  CuPc thin film  grown on ITO coated glass at room temperature for three thicknesses(100nm, 200nm and 400nm).

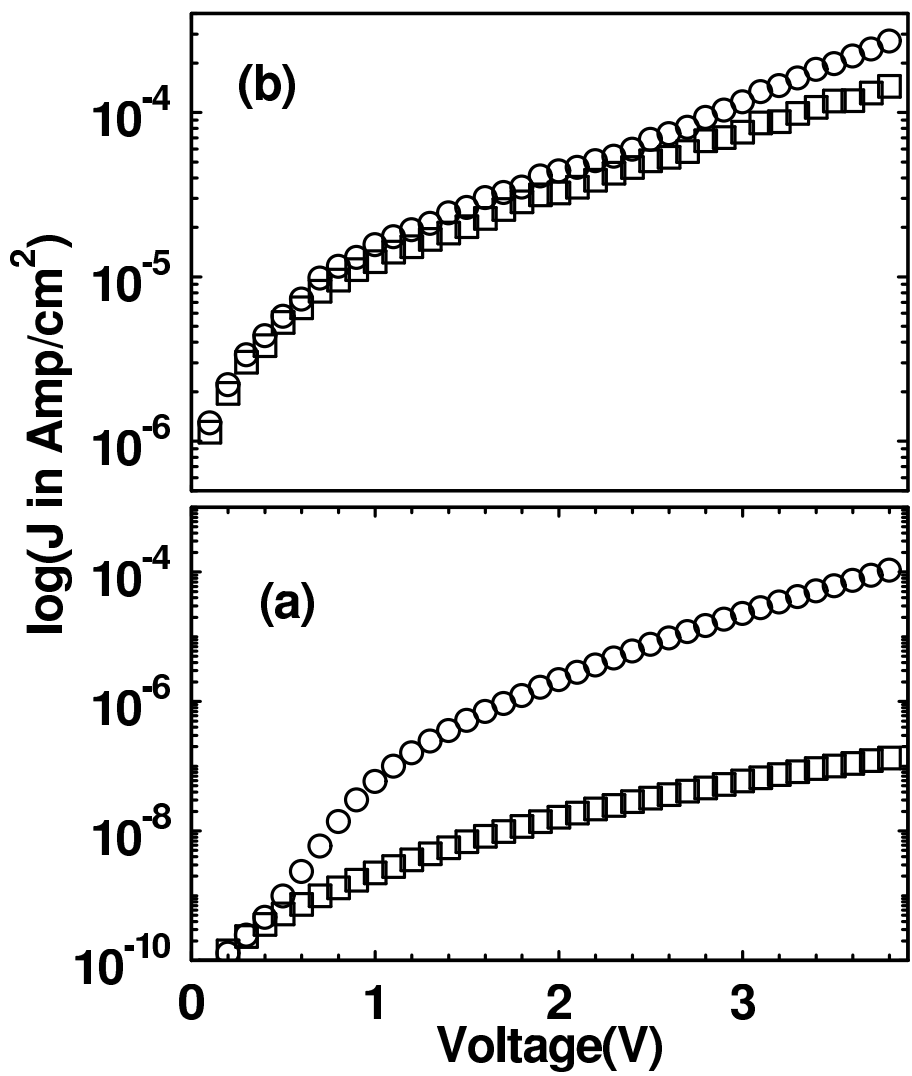
\item[Fig.2.] Room temperature J-V characteristics for single layer hole only CuPc device with thickness of 100nm for two structures.  Empty circles represent the data when  ITO is positively biased and empty squares represent the data when ITO is negatively biased, in (a)ITO/CuPc/Cu and (b)ITO/CuPc/Al.  

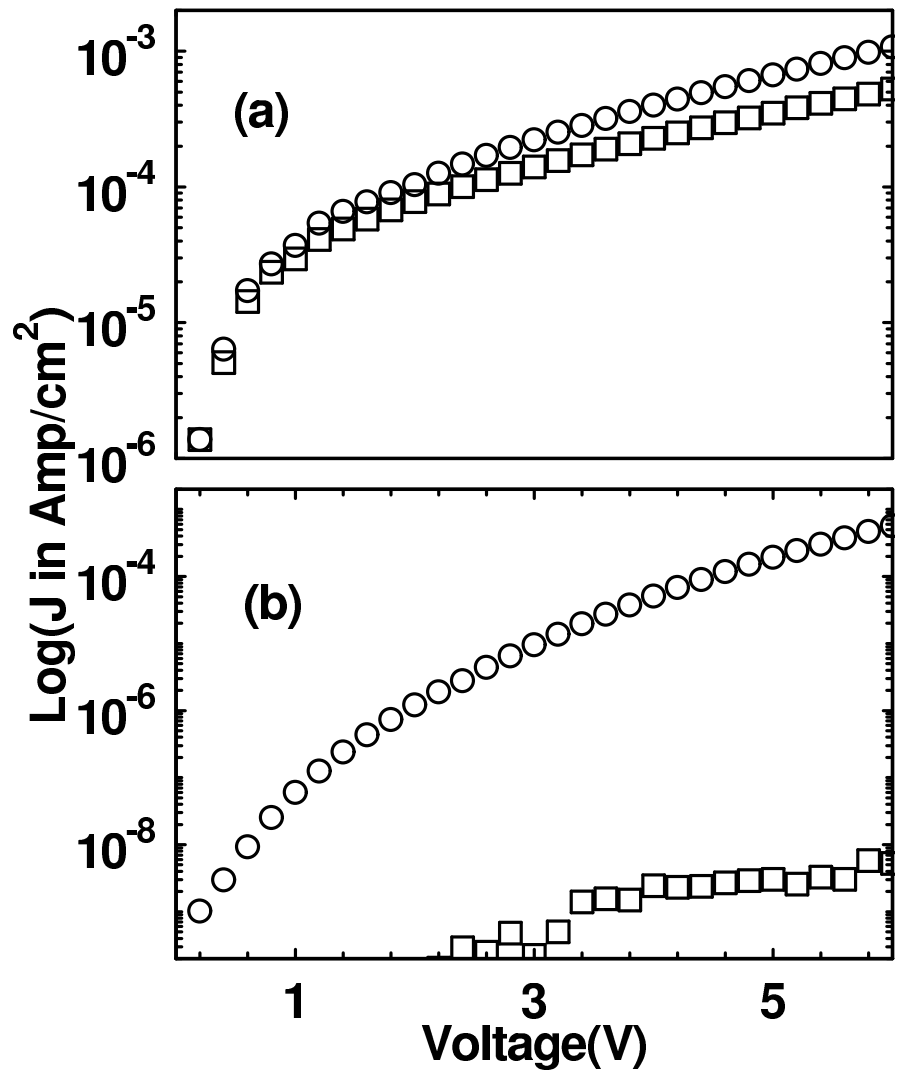
\item[Fig.3.] Room temperature J-V characteristics for single layer hole only CuPc device with thickness of 200nm for two structures.  Empty circles represent the data when  ITO is positively biased and empty squares represent the data when ITO is negatively biased, in (a)ITO/CuPc/Cu and (b)ITO/CuPc/Al. 

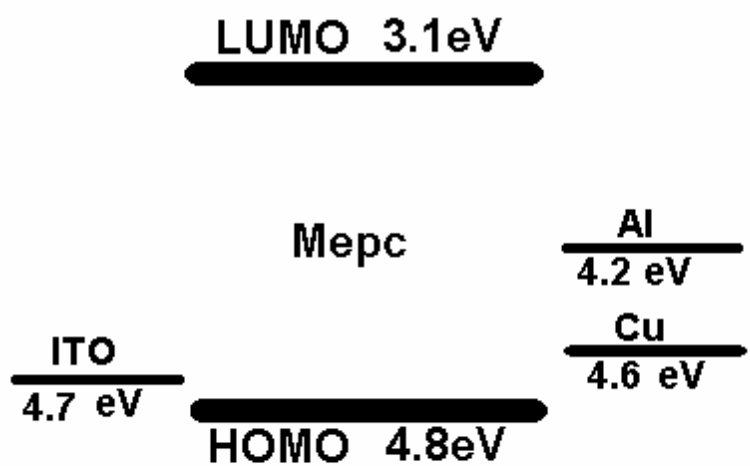
\item[Fig.4.] Schematic illustration of the energy level diagram for the MePc based devices in Fig.2 and Fig.3. 

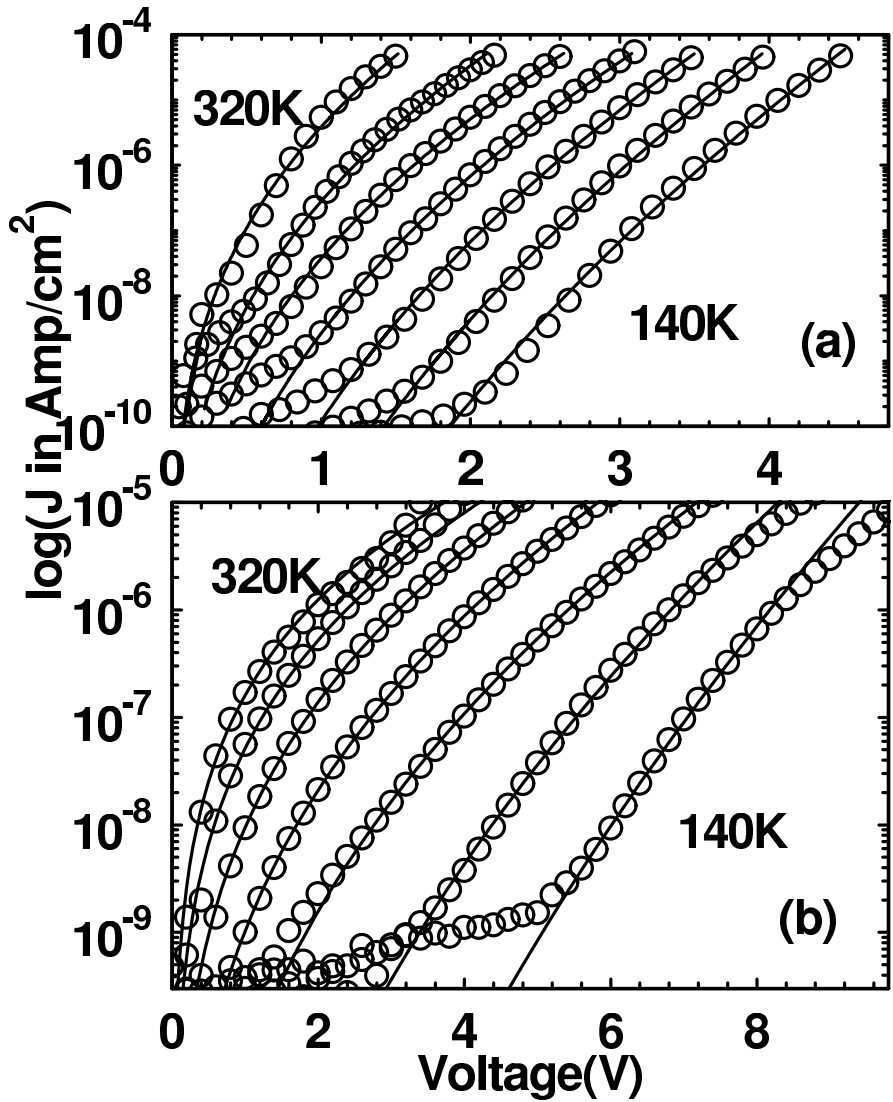
\item[Fig.5.]    J-V characteristics of    ITO/CuPc/Al  at different temperature staring at 320K and then at the interval of 30K for (a) 100nm CuPc and (b) 200nm CuPc.  Empty circles show experimental data and solid lines represent the  simulated data. 

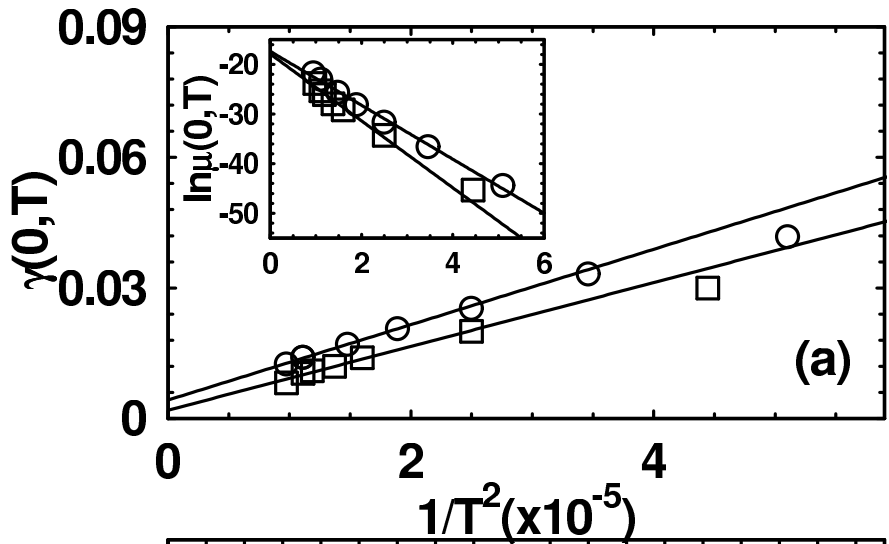
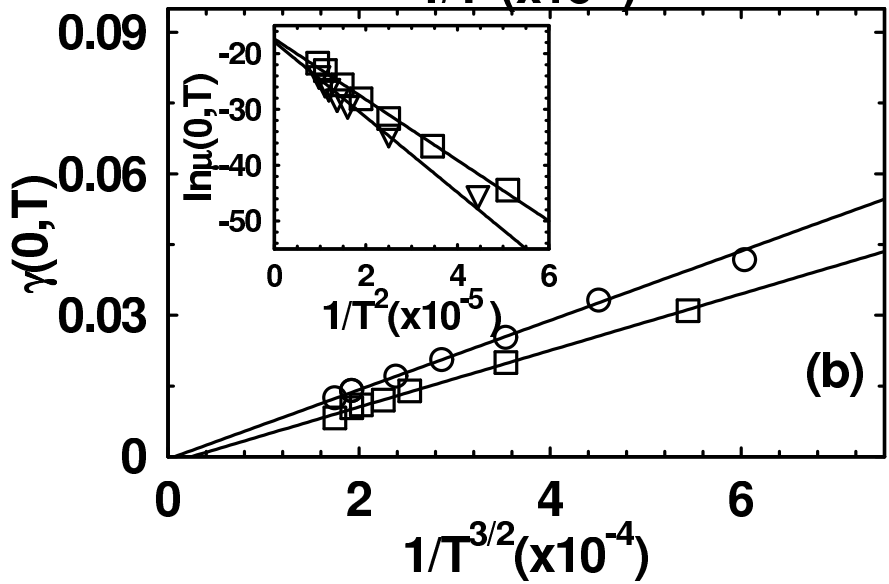
\item[Fig.6.] Temperature dependence of  the field activation parameter $\gamma(0,T)$(in (cm/V)$^{1/2}$)   for CuPc(empty circles) and ZnPc(empty squares)  obtained from the  fit to the experimental data given in Fig.5 using  (a) UGDM and (b)CGDM. Insets show zero field mobility $\mu(0,T)$(in cm$^2$/Vs) in these two  cases.

\end{description}

\end{document}